# Diagnostic Imaging for Damage Detection in Plates Based on Topological Acoustic (TA) Sensing Technique

Bo Hu[1, 2], Tribikram Kundu[1-4,*], Pierre A. Deymier[1, 4], Keith Runge[1, 4]

1. New Frontiers of Sound Science and Technology Center, University of Arizona, Tucson, AZ 85721, USA.

2. Department of Aerospace and Mechanical Engineering, University of Arizona, Tucson, AZ 85721, USA.

3. Department of Civil and Architectural Engineering and Mechanics, University of Arizona, Tucson, AZ 85721, USA.

4. Department of Materials Science and Engineering, University of Arizona, Tucson, AZ 85721, USA.

[*]Corresponding author: tkundu@arizona.edu

**Abstract:** Traditional structural damage detection methods in aerospace applications face challenges in accuracy and sensitivity, often necessitating multiple sensors to evaluate various measurement paths between the reference and defective states. However, the recently developed topological acoustic (TA) sensing technique can capture shifts in the geometric phase of an acoustic field, enabling the detection of even minor perturbations in the supporting medium. In this study, a diagnostic imaging method for damage detection in plate structures based on the TA sensing technique is presented. The method extracts the geometric phase shift index (GPS-I) from the Lamb wave response signals to indicate the location of the damage. Using Abaqus/CAE, a finite element model of the plate was established to simulate the Lamb wave response signals, which were then used to validate the feasibility of the proposed method. The results indicate that this technique enables rapid and precise identification of damage and its location within the plate structure, requiring response signals from only a few points on the damaged plate, and it is reference-free.

# 1. Introduction

The presence and progressive accumulation of structural damage in aerospace structures pose significant threats to both safety and structural integrity, potentially leading to catastrophic consequences. Therefore, it is crucial to prioritize the development of intelligent structures equipped with self-sensing capabilities, which hinges on the refinement and integration of structural health monitoring (SHM) techniques [1-2].

To address challenges in aerostructures, researchers focus on two key areas: acoustic source localization (ASL) and damage identification and localization (DIL). ASL is a real-time monitoring technique that accurately identifies the location of acoustic events, such as collisions or crack propagation, in aircraft wings and fuselages, enabling swift responses to safety risks. In contrast, DIL proactively detects and assesses structural damage, aiming to identify and accurately locate issues for effective maintenance. DIL often incorporates advanced algorithms and technologies like machine learning to enhance detection. While ASL focuses on immediate monitoring, DIL emphasizes periodic inspections and long-term maintenance. Together, they provide a comprehensive approach to structural health monitoring, improving safety and reducing operational costs [3-6].

The techniques for ASL include time-of-arrival (TOA), time difference of arrival (TDOA), frequency difference of arrival (FDOA), and Doppler shift, all of which depend on a network of sensor nodes with known locations [7-19]. Comparatively, TOA and TDOA methods provide higher positioning accuracy and require only one channel per sensor node for measurements, thereby minimizing the load on individual sensor nodes, making them among the most commonly used techniques. However, both methods necessitate accurate parameter determination, which is frequently affected by environmental noise. Alternately, some studies utilize the energy of recorded signals to identify the acoustic source without depending on TOA or TDOA data. One prominent technique is beamforming, a delay-and-sum approach that selectively enhances or reduces signals in a specific direction by integrating inputs from multiple sensors. Its goal is to precisely locate acoustic sources while delivering directional information, but this method can incur a substantial computational burden.

In the field of damage identification and localization (DIL), the time reversal technique is a promising and extensively researched method. This approach utilizes the time-reversibility of waves to detect and localize the damage [20-23]. Additionally, inspired by principles from radar and medical imaging, phased array methods have also been applied for damage identification and localization in plates. Phased arrays consist of actuators or sensors with varying spatial distributions, allowing for control of the effective propagation and sensing directions through algorithmic adjustments of their relative phases. Echo signals from the damaged area can be analyzed to provide indicators of damage [24-27]. However, due to stringent spatial resolution requirements, phased array methods often necessitate a large number of sensors and relatively high excitation frequencies to achieve the desired accuracy and performance in applications such as target localization and tracking.

However, damage imaging is typically superior to conventional defect detection because it not only identifies existing flaws but also provides more detailed information, such as their shape, size, location, and distribution. Through imaging techniques, data can be obtained across the entire surface or volume of the material, facilitating comprehensive assessments of structures or components. Moreover, imaging techniques often have higher spatial resolution and sensitivity, enabling detection of even minor flaws or changes, thereby providing more accurate assessments and predictions to prevent potential damage or failure. Consequently, damage imaging is considered a more advanced and comprehensive defect detection method, offering engineers and scientists more information to guide repair and maintenance efforts.

The recently developed nonlinear ultrasonic technique, Sideband Peak Count (SPC), together with its derivatives—including the Sideband Peak Count-Index (SPC-I) and the Sideband Intensity Index (SII)—exhibits enhanced robustness and sensitivity in detecting various types of early-stage or small-scale damage, such as micro-fatigue crack initiation and incipient corrosion, and other forms of material degradation [2, 28]. Previously, we proposed a diagnostic imaging method based on the SPC technique for detecting damage in plate structures. The results demonstrate that this approach can accurately detect and localize damage, offering a promising solution to enhance structural health monitoring across a range of engineering applications [29]. Another recent investigation by the New Frontiers of Sound Science and Technology Center (NewFoS) found that even minor alterations in the supporting medium of acoustic fields can induce significant shifts in geometric phase [30-34]. This novel method called "topological acoustic sensing" utilizes changes in geometric phase of acoustic fields to detect various types of defects in structures. Currently, the research on using topological acoustic sensing for defect detection is in its early stages. Although this technology shows promise in defect detection and localization, it is still in its nascent stage and has several limitations. These limitations include the need to deploy a considerable number of sensors to probe the acoustic fields and subsequently construct complex vector representations in a multidimensional Hilbert space and its inability to achieve baseline-free damage detection.

This study aims to develop a diagnostic imaging method for damage detection using acoustic sensing based on geometric phase, addressing existing challenges and enhancing a previously proposed method based on the SPC technique through the incorporation of topological acoustic sensing technology. The objectives of this research include two aspects: (1) to apply topological acoustic sensing techniques for detecting and monitoring damages in plate-like structures, along with developing corresponding algorithms for damage imaging; and (2) to validate the effectiveness and applicability of the proposed method through numerical modeling of Lamb wave propagation in damaged plate-like structures.

## 2. Theory and algorithm

Topological acoustic sensing captures alterations in the geometric phase of an acoustic field (linear and/or nonlinear), presenting them as changes in its vectorial representation within a multidimensional Hilbert space resulting from a disturbance. Previous research utilizing topological acoustic sensing has shown that even minor modifications in the medium supporting an acoustic field can induce significant alterations in the geometric phase. In cases where the topology of the manifold in the multidimensional space, spanned by the vectorial representation of acoustic fields, displays sharp topological features such as twists, slight alterations in the supporting medium may lead to pronounced shifts in geometric phase. Monitoring variations around such features enhances the geometric phase's sensitivity to minor disturbances.

Several studies have been conducted on topological acoustic sensing for defect detection. The fundamental principles of topological acoustic sensing and geometric phase change are briefly summarized below.

### 2.1 Topological acoustic sensing and geometric phase shift

Let an acoustic source be placed at a point on a geometric entity, from which acoustic signals (such as displacements and velocities) are captured at a random set of $N$ points. Consequently, time series data can be acquired at those $N$ locations within the acoustic field. A fast Fourier transform (FFT) is then applied to each of these N time series to extract the complex amplitudes in the spectral domain. At a specific frequency, these $N$ complex amplitudes can be depicted as normalized state vectors within an N-dimensional complex Hilbert space:

$$|\psi\rangle = \begin{pmatrix} \psi_1 \\ \psi_2 \\ \vdots \\ \psi_N \end{pmatrix} = \begin{pmatrix} |\psi_1|e^{j\phi_1} \\ |\psi_2|e^{j\phi_2} \\ \vdots \\ |\psi_N|e^{j\phi_N} \end{pmatrix} \qquad (1)$$

Where $\psi_i = |\psi_i|e^{j\phi_i}$ ($i = 1, 2, \ldots, N$) is the complex amplitudes at each receiving point, while $|\psi_i|$ and $\phi_i$ are magnitude and phase at each receiving point, respectively. Assuming that the state vector $|\psi\rangle$ in the Hilbert space represented by Eq. (1) corresponds to a defect-free state. Then, after measurement, it is possible to obtain another state vector in a different Hilbert space corresponding to a defected state, which we can denote as $|\psi'\rangle$.

Therefore, at a single given frequency, the angle between the Hilbert space vectors representing the acoustic field at $N$ locations in the undamaged and damaged systems corresponds to the geometric phase change of the acoustic wave. This angle, or the single geometric phase change at the given frequency, can be obtained through the dot product of these two state vectors and can be expressed as [31-32]:

$$\Delta\varphi = arcos(Re(\frac{\langle\psi|\psi'\rangle}{\|\psi\|\|\psi'\|})) \qquad (2)$$

Where $\langle\psi|\psi'\rangle = \sum_i^N \psi_i^* \psi_i'$ is the inner product between Hilbert space vector $|\psi\rangle$ and $|\psi'\rangle$; $\|\psi\| = \langle\psi|\psi\rangle^{\frac{1}{2}}$ and $\|\psi'\| = \langle\psi'|\psi'\rangle^{\frac{1}{2}}$ are respectively the norm of the vector $|\psi\rangle$ and $|\psi'\rangle$; Re stands for the real part of a complex quantity.

The geometric phase shift plots for each frequency component can be represented as a function of frequency. While the geometric phase shift-index (GPS-I) can be defined as the average value of $\Delta\varphi(\omega)$ within a specific frequency range:

$$\text{GPS-I} = \frac{\int_{\omega_{min}}^{\omega_{max}} \Delta\varphi(\omega)d\omega}{\omega_{max} - \omega_{min}} \qquad (3)$$

where GPS-I is the geometric phase shift-index; $\Delta\varphi(\omega)$ is the geometric phase shift that corresponds to frequency $\omega$; $\omega_{max}$ and $\omega_{min}$ are the upper and lower frequency limits, respectively.

**2.2 Damage imaging algorithm**

Let us consider a square plate with edge length $L$. The plate has an actuator positioned at its center and sensors located at each of its four corners, as illustrated in Fig. 1. Then, the length of the $i$-th propagation path for sensors located at the corners can be expressed as: $d_i = \sqrt{[2(i-1)+1]^2 + 1} \cdot \frac{L}{2}$ ($i$ = 1, 2, 3…). Then one can estimate the arrival time of the propagating wave for each path and each segment of the entire signal. Although an infinite number of paths exists, only the first three propagation paths are considered in this study.

Meanwhile, we selected the minimum number of sensors, which is four. To effectively capture the geometric phase shift, the sensor arrangement in damage detection should adhere to symmetry principles as much as possible. Therefore, two highly symmetric arrangements were considered: one with sensors placed at the four corner points of the plate, and the other with sensors placed at the midpoints of each edge, as shown in the lower-left corner of Fig. 1.

Placing sensors at the four corners of the plate maximizes the use of diagonal lengths, enabling longer wave propagation paths and more comprehensive structural coverage. In contrast, the midpoint arrangement tends to concentrate propagation paths near the center of the plate, potentially leaving edge areas insufficiently monitored. Moreover, from a waveform identification perspective, it is desirable to avoid overlap between wave packets corresponding to different propagation paths, including direct transmission between the actuator and sensor as well as paths involving one or more boundary reflections. In the corner configuration, the propagation distances vary significantly, facilitating time-domain separation of wave packets. However, in the midpoint configuration, the propagation paths are more uniform in length, increasing the likelihood of wave packet superposition and reducing signal distinguishability. Two different damage locations are considered. If the center point of the plate is taken as the origin of the coordinates, the coordinates of these two damage points are (60, 100) and (60, -40), each represented by a circular hole with a diameter of 6 mm, as shown in the enlarged view of the plate in the upper right corner of Fig. 1.

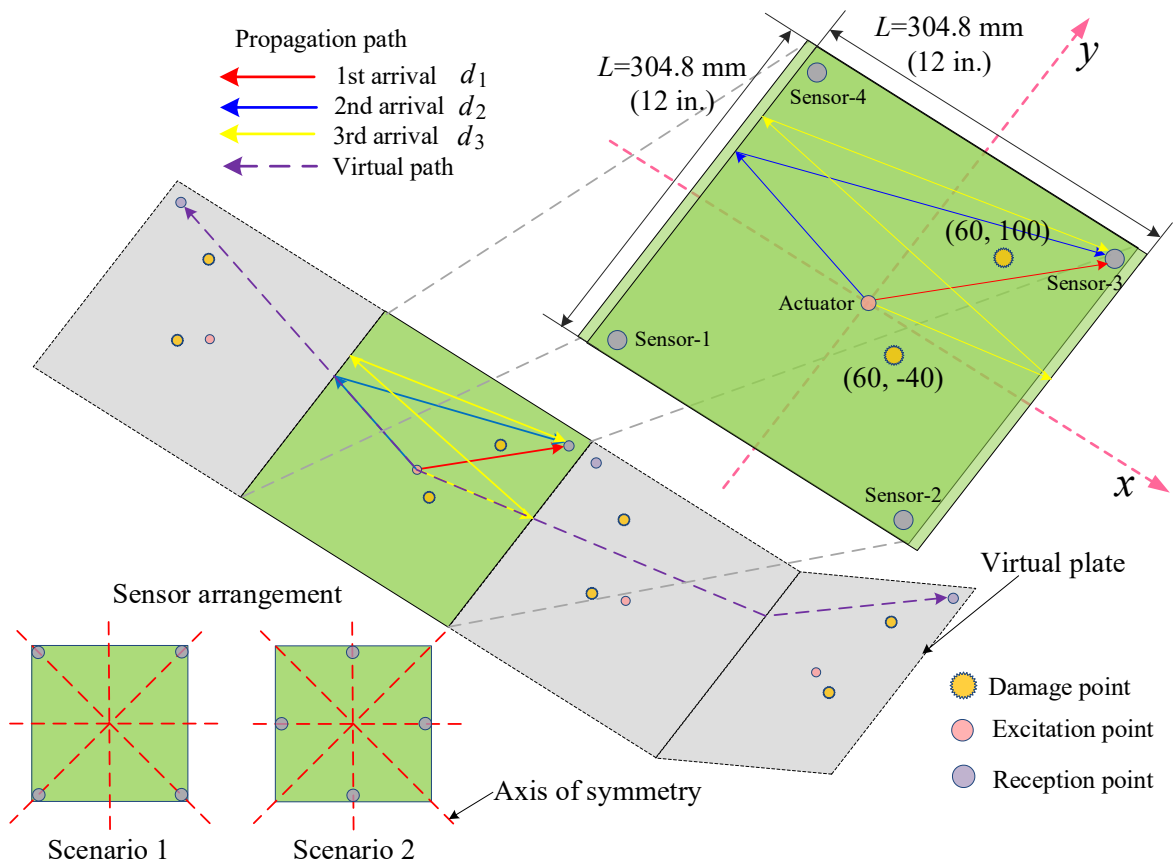

Fig. 1. Schematic of the aluminum plate and the wave propagation paths.

This is equivalent to transforming the second to $N$-th propagation paths into straight lines through mirroring. As a result, if we segment and extract the received signals from each sensor using the method mentioned above, it is as if we have $(N-1)$ virtual sensor arrays receiving signals mirrored after the original ones. Additionally, by including the signal received from the first propagation path, which is already a straight line, a $N$-dimensional Hilbert space can be constructed. It should be noted that the $N$-dimensional Hilbert space here is constructed from only one signal received by only one sensor by separating different segments of that signal. It is important to emphasize that the dimension of the Hilbert space is that the number of paths used, which correspond to the number of virtual sensors along with those actual physical sensors.

If we only consider the uniformity of the damage occurrence probability at each point, without emphasizing the actual probability values, then the probability of damage occurrence at each point can be defined as:

$$P(x_p, y_q) = \exp(\text{GPS-I}_{conv}(x_p, y_q)) / \sum_q \sum_p \exp(\text{GPS-I}_{conv}(x_p, y_q)) \quad (4)$$

Where $P(x_p, y_q)$ represents the probability of damage occurring at a specific point on the plate, and $(x_p, y_q)$ denotes the coordinates of that point; $\text{GPS-I}_{conv}(x_i, y_j)$ denotes the convolution result of the GPS-I distribution with a 2D Gaussian kernel $G(u,v) = \frac{1}{2\pi\sigma^2}\exp(-\frac{u^2+v^2}{\sigma^2})$, as follows:

$$\text{GPS-I}_{conv}(x,y) = (\text{GPS-I} * G)(x,y) = \sum_v \sum_u \text{GPS-I}(x-u, y-v) G(u,v) \quad (5)$$

It is reasonable to assume that the closer the point $(x_p, y_q)$ is to the damaged point, the larger is the value of $P(x_p, y_q)$. It can be approximated as a two-dimensional normal distribution:

$$P(x_p, y_q) = \frac{1}{2\pi\sigma_x\sigma_y}\exp\left(-\frac{1}{2}\left[\frac{(x_p-u_x)^2}{\sigma_x^2} + \frac{(y_q-u_y)^2}{\sigma_y^2}\right]\right) \quad (6)$$

Where $u_x$ and $u_y$ are the means of the distribution for the $x$ and $y$ variables, respectively; $\sigma_x$ and $\sigma_y$ are the standard deviations of the distribution for the $x$ and $y$ variables, respectively.

Since, Eq. (4) and (6) represent the same probability distribution, the coefficients in Eq. (6) can be estimated using the distribution characteristics of GPS-I. It should be noted that the GPS-I used in the above equation is for a specific frequency. However, considering a set of frequencies $\omega_r, (r = 1, 2, \ldots, M)$, we can obtain a series of different probability distributions $P_r(x_p, y_q), (r =$

$1, 2, \ldots, M$). The most probable damage location is the region with the highest probability value when the $N$ probability distributions are multiplied together. Therefore, we can define the damage index as follows.

$$DI(x_p, y_q) = \left(\prod_{r=1}^{M} P_r(x_p, y_q) - \min\left(\prod_{r=1}^{M} P_r(x_p, y_q)\right)\right) \Big/ \left(\max\left(\prod_{r=1}^{M} P_r(x_p, y_q)\right) - \min\left(\prod_{r=1}^{M} P_r(x_p, y_q)\right)\right) \quad (7)$$

where $DI(x_p, y_q)$ represents the damage index at point $(x_p, y_q)$, $P_r(x_p, y_q)$ represents the probability at point $(x_p, y_q)$ corresponding to $r$-th frequency $\omega_r$.

## 3. Numerical simulation

In a plate, Lamb waves comprise a mix of symmetric and anti-symmetric modes, with their dispersion relations often calculated using Rayleigh-Lamb frequency equations [35].

$$\frac{\tan(qh)}{\tan(ph)} = -\frac{k^2 pq}{(q^2 - k^2)^2} \quad \text{for symmetric modes} \quad (8a)$$

$$\frac{\tan(qh)}{\tan(ph)} = -\frac{(q^2 - k^2)^2}{k^2 pq} \quad \text{for antisymmetric modes} \quad (8b)$$

Where $h=$ plate thickness, $k=$ wave number; $p$ and $q$ are the characteristic values that can be defined as follows:

$$p^2 = \frac{\rho}{\lambda + 2\mu}\omega^2 - k^2, \qquad q^2 = \frac{\rho}{\mu}\omega^2 - k^2 \quad (9)$$

Where $\omega =$ wave frequency, $\lambda$ and $\mu$ are Lame constants.

The Rayleigh-Lamb frequency equations typically lack analytical solutions, hence numerical methods are required for solving them. By solving these equations, dispersion curves can be obtained. The dispersion curves for group velocity and phase velocity of an isotropic aluminum plate are depicted in Figs. 2 (a) and (b), respectively.

In the received wave response, multiple symmetric and asymmetric wave modes can superpose. As the frequency increases, the number of coexisting wave modes also increases, as shown in Fig. 2. Therefore, a high-frequency excitation signal would introduce numerous wave modes into the response, complicating the analysis of the wave behavior.

In this study, the excitation signal is a Hanning-windowed tone burst $S(t)$, and it can be expressed as follows [36].

$$S(t) = \begin{cases} \frac{1}{2}\left[1 - \cos\left(\frac{2\pi f t}{n_c}\right)\right]\sin(2\pi f t) & \text{if } 0 \leq t \leq \frac{n_c}{f} \\ 0 & \text{Otherwise} \end{cases} \quad (10)$$

Where $f$ is the excitation frequency; and $n_c$ is the number of cycles. In this study, $f$ was 215 kHz and $n_c$ was equal to 3.5. The time-domain and frequency-domain characteristics of the excitation signal are shown in Figs. 3 (a) and (b), respectively.

Herein, it needs to be emphasized that the main reason for selecting a narrow frequency range is to avoid signal overlap. As shown in Fig. 2, the theoretical dispersion curves for an aluminum plate illustrate that, in a broader frequency range, the dispersion effect during wave propagation causes the wave speeds of different frequency components to separate, resulting in overlap between different wave packets. This overlap complicates signal identification and analysis. Therefore, by using a narrower frequency range, we ensure better time-frequency consistency of the wave packets, which enhances the accuracy and reliability of damage detection.

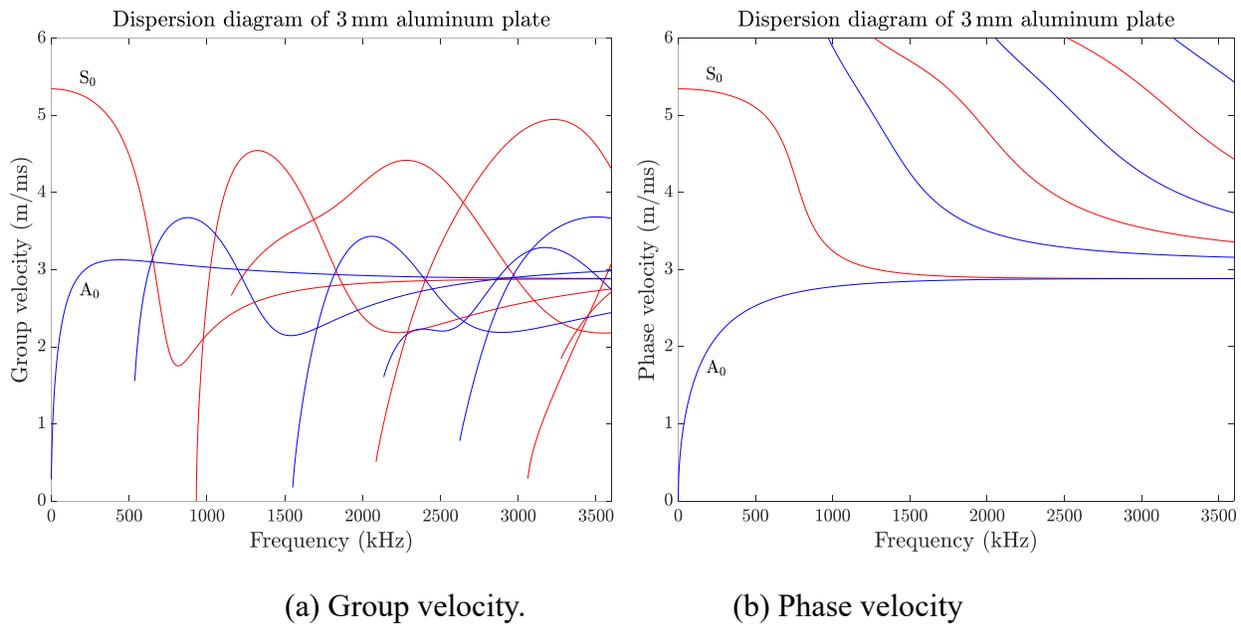

(a) Group velocity.  (b) Phase velocity

Fig. 2. Theoretical dispersion curves for an aluminum plate of thickness (3 mm): (a) Group velocity; and (b) Phase velocity.

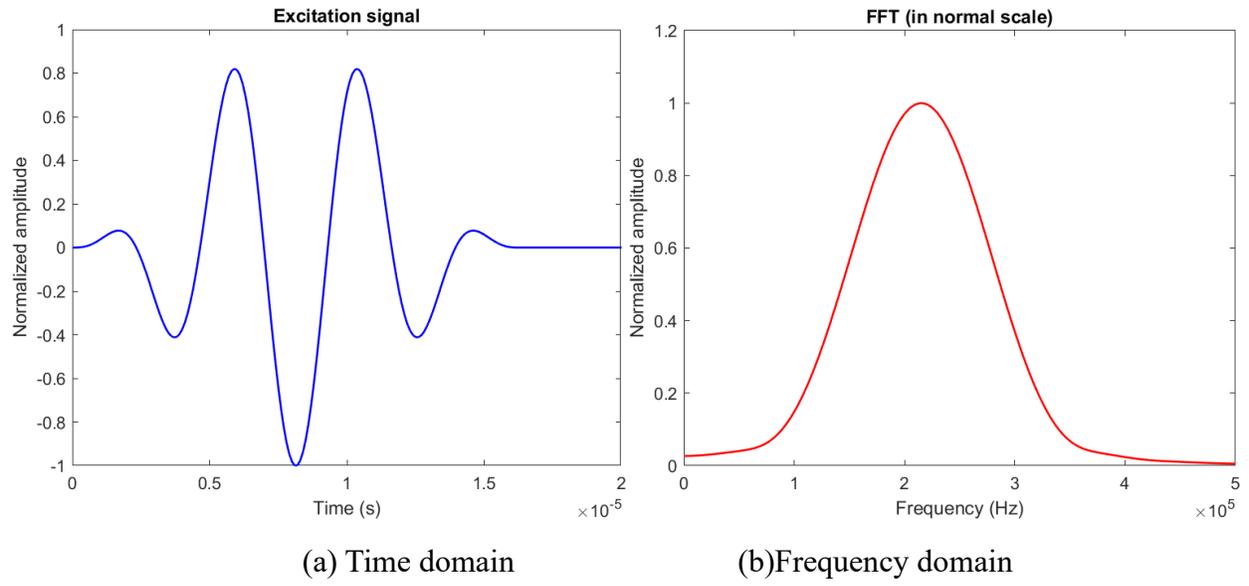

(a) Time domain          (b) Frequency domain

Fig. 3. Excitation signal in: (a) Time domain; and (b) Frequency domain.

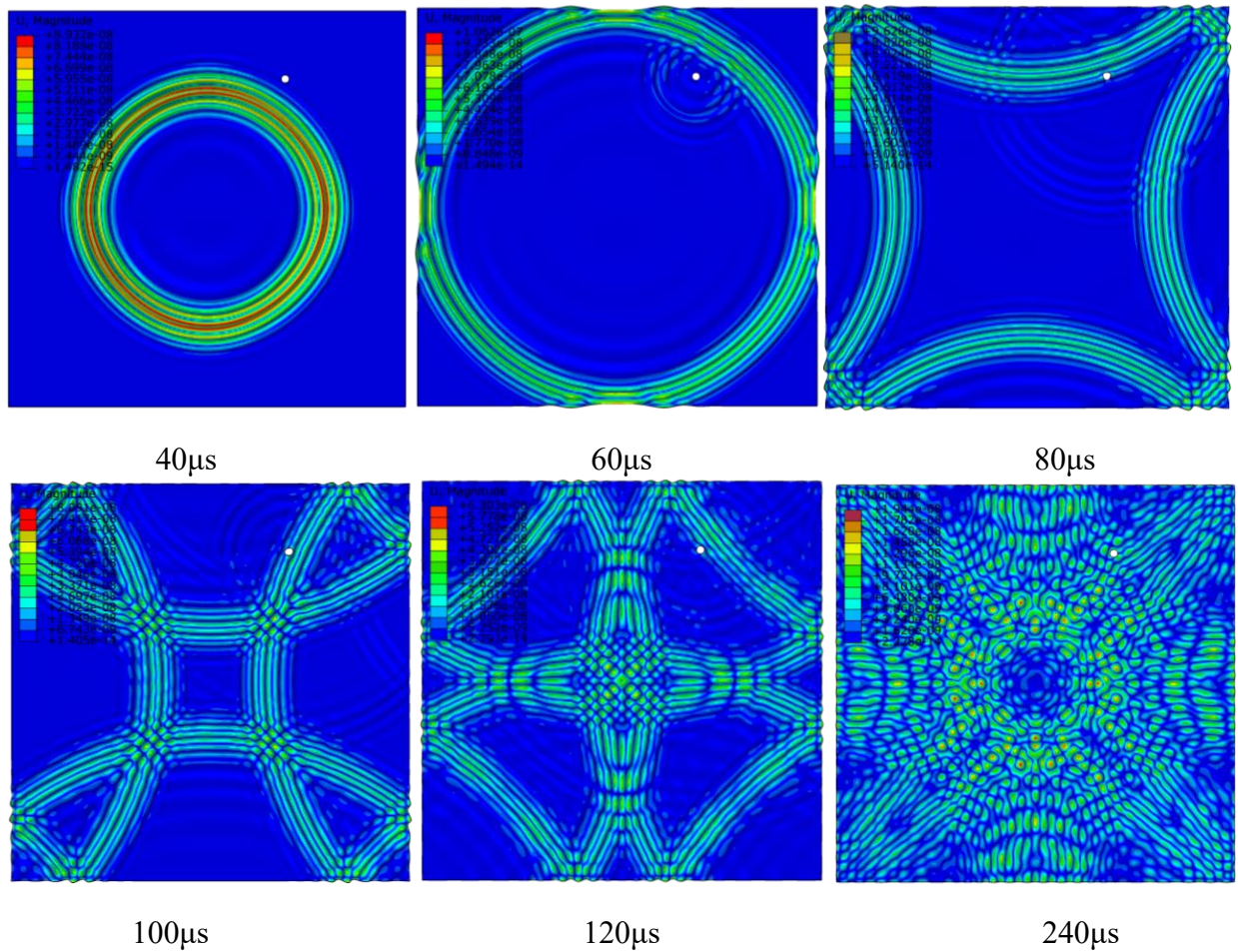

40μs          60μs          80μs

100μs          120μs          240μs

Fig. 4. Wave motion observations in a damaged aluminum plate at different time instances for

the damage location coordinates at (60, 100). (Scale factor $= 1 \times 10^8$)

Using the ABAQUS finite element package, aluminum plate models were established with S4 elements, as shown in Fig. 4. The damage is indicated as a white dot in the upper right quadrant of the plate, representing a defect in the form of a circular hole with a diameter of 6 mm, located at the coordinates (60, 100). Fig. 4 also illustrates the propagation of Lamb waves in a damaged plate at 40 μs to 240 μs. The response signals captured by the four sensor locations are shown in Fig. 5 (a). Three segments of sub-signals were extracted from each response signal, corresponding to the first three different propagation paths, as illustrated in Figs. 5 (b) to (d).

## 4. Topological acoustic sensing-based analysis

### 4.1 Time Signal Processing

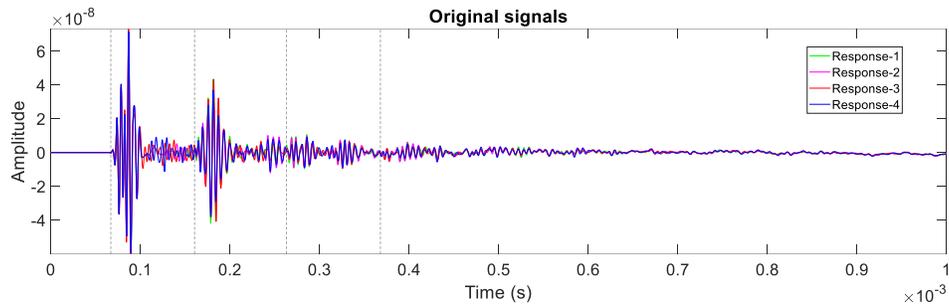

(a) Original signal

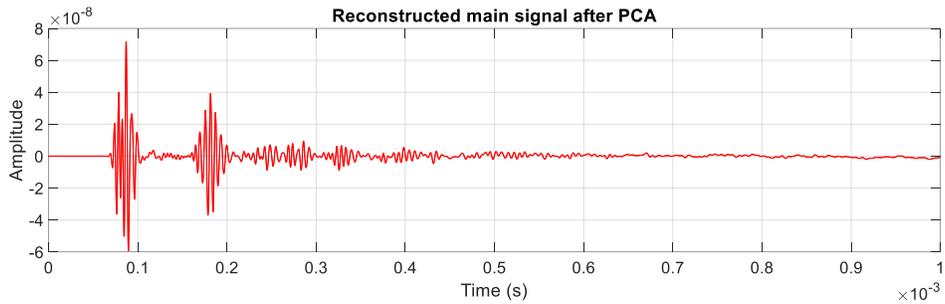

(b) Reconstructed signal

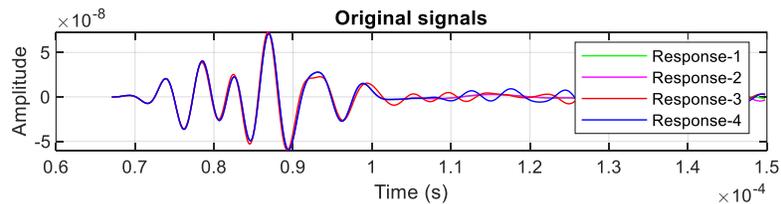

(c) 1st wave packet

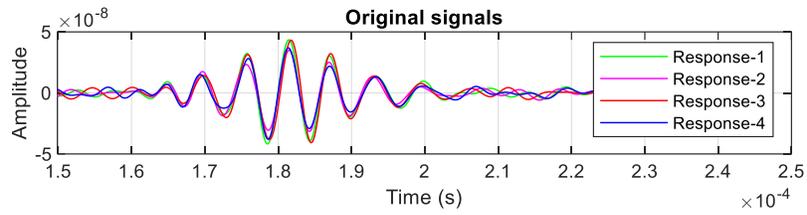

(d) 2nd wave packet

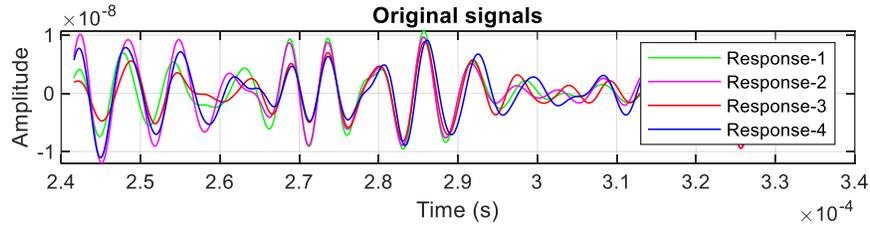

(e) 3rd wave packet

Fig. 5. Response signals: (a) Original signal; (b) Reconstructed signal after principal component analysis; (c) 1st wave packet; (d) 2nd wave packet; and (e) 3rd wave packet.

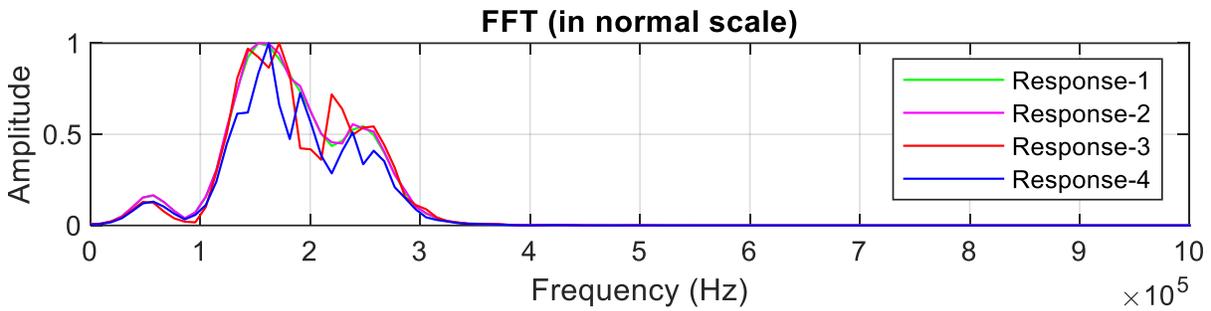

(a) 1st wave packet

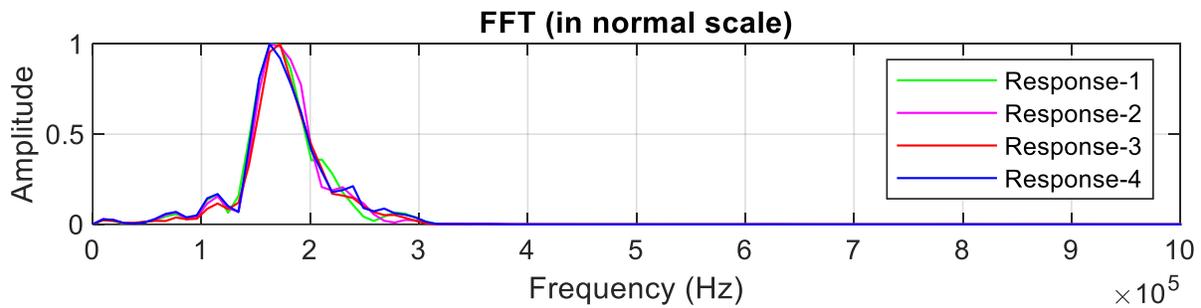

(b) 2nd wave packet

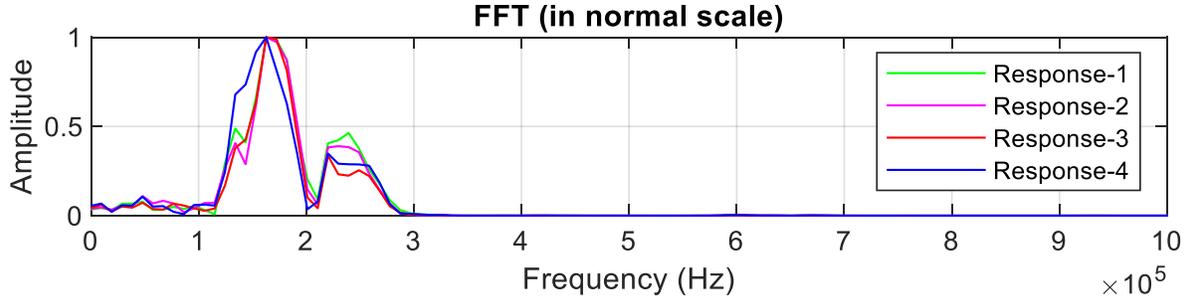

(c) 3rd wave packet

Fig. 6. The spectrum of three wave packets displayed in normalized scale: (a) 1st wave packet, (b) 2nd wave packet, and (c) 3rd wave packet.

The segmented response signals were subjected to Fast Fourier transform (FFT), resulting in the spectrum of each received signal corresponding to each propagation path, as shown in Fig. 6. In general, TA sensing requires selecting various sensor positions to collect signals, which are then processed through Fast Fourier Transform (FFT) to construct an N-dimensional complex Hilbert space. However, in this study, we utilize the signal value from a single measurement point to construct the complex Hilbert space. Although the signal is collected from one point, the complete signal is segmented based on the arrival times for different propagation paths of the three Lamb wave packets. The first wave packet corresponds to the signal recoded at the actual sensor, while the second and third wave packets can be regarded as signals recorded by virtual sensors in the spatial domain, as the propagation paths illustrated in Fig.1. Thus, the segmented signals collected by the actual and virtual sensing points correspond to the three different propagation paths and can be used to form the complex Hilbert space.

### 4.2 Geometric phase shift

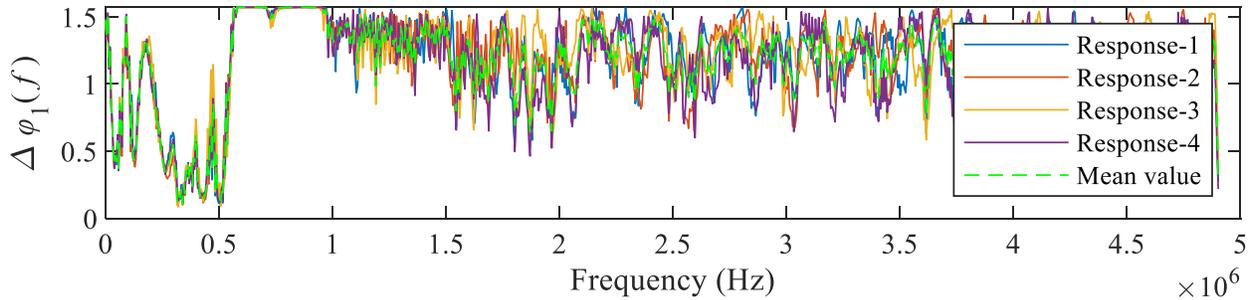

(a) Phase shift

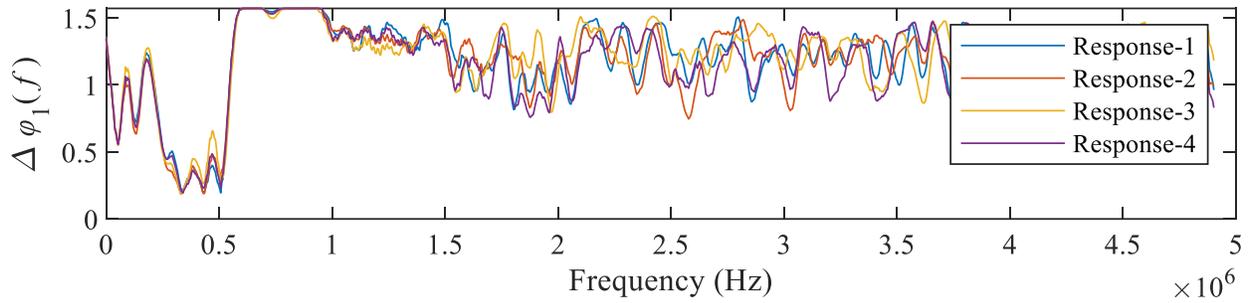

(b) Smooth trended phase shift

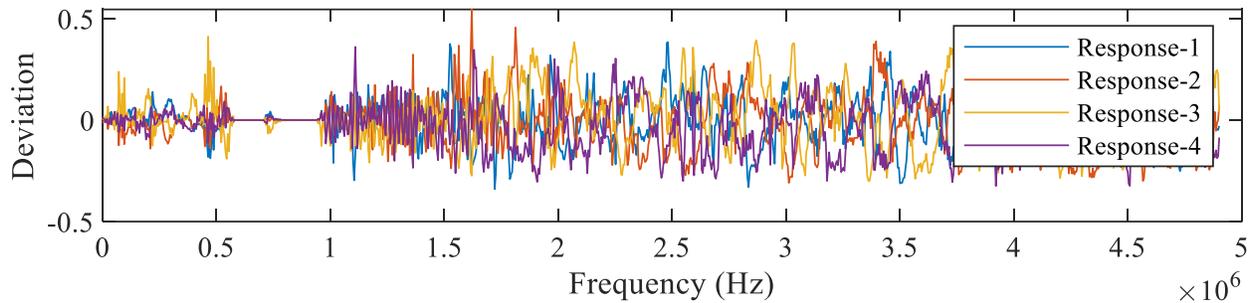

(c) Deviation from mean

Fig. 7. Phase shift component and deviation for 1st wave package: (a) Phase shift; (b) Smooth trended phase shift; and (c) Deviation from mean.

Due to the geometric symmetry of the plate and the sensor positions, the signals collected by the four sensors should be identical when the plate is undamaged. However, when there is damage on the plate, additional reflected signals will be introduced due to scattering from the defect, resulting in variations in the signals received by each sensor at different times. Specifically, the signals received by the four sensors at different locations can be viewed as the same primary signal (including the first wave signal and boundary reflections) with the addition of weaker reflections caused by the damage. Therefore, these signals exhibit similarities in their characteristics. By leveraging this similarity, Principal Component Analysis (PCA) can be applied to extract the primary signal, which can then serve as a baseline, as shown in Fig.5 (b).

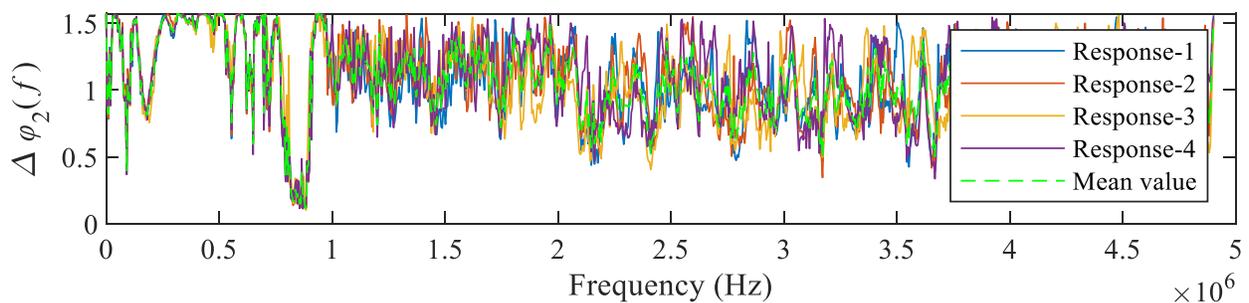

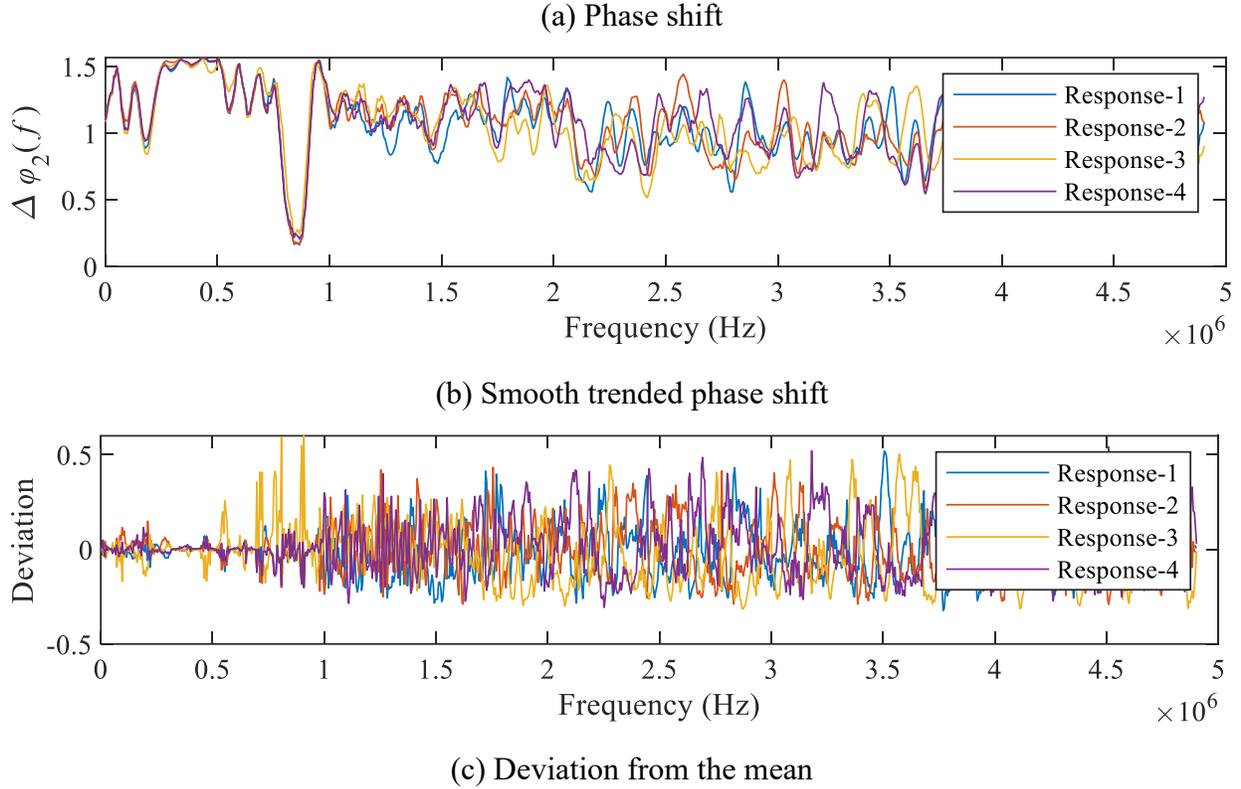

Fig. 8. Phase shift component and deviation for 2nd wave packet: (a) Phase shift; (b) Smooth trended phase shift; and (c) Deviation from the mean.

From Eq. 2, the total phase shifts ($\Delta\varphi$) under two different conditions can be computed. However, by retaining only the signal value from one propagation path and setting the signal values from the other two paths to zero, the phase shift components ($\Delta\varphi_i$) corresponding to each Lamb wave propagation path can still be calculated using the same equation. Therefore, Eq. 2 can be expressed as: $\Delta\varphi_i = arcos(Re(\frac{\langle\psi_i|\psi'_i\rangle}{\|\psi_i\|\|\psi'_i\|}))$, $i$=1, 2, and 3. This approach enables an independent analysis of the contribution of each propagation path to the phase shift.

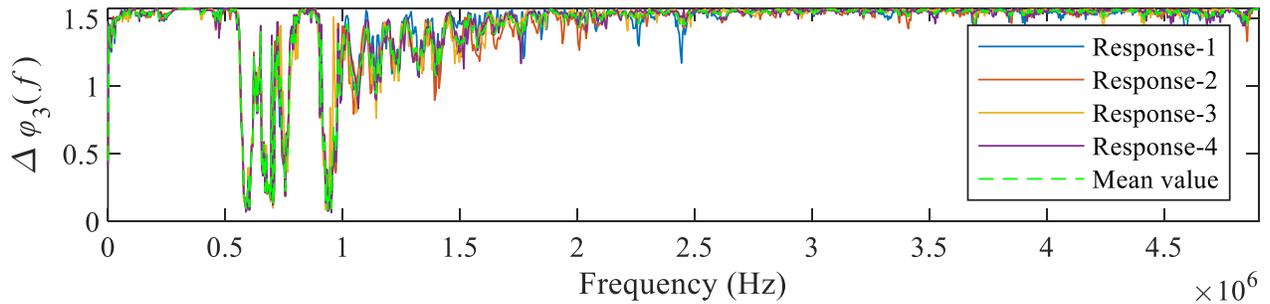

(a) Phase shift

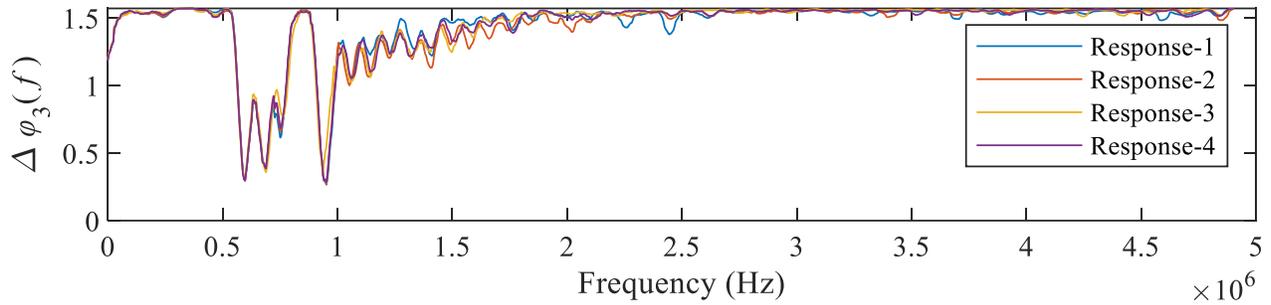

(b) Smooth trended phase shift

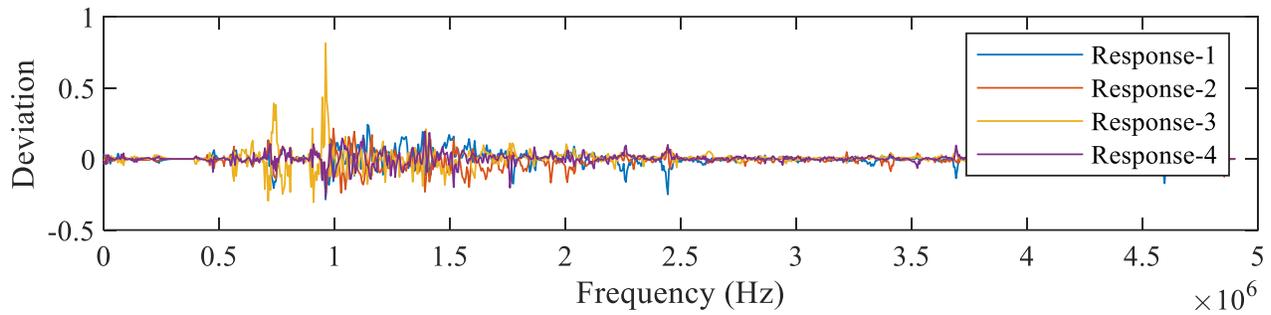

(c) Deviation from mean

Fig. 9. Phase shift component and deviation for the 3rd wave package: (a) Phase shift; (b) Smooth trended phase shift; and (c) Deviation from the mean.

Since the phase shift components of the three wave packets exhibit consistent and similar patterns, they will be introduced and discussed together here. The phase shift components corresponding to the first three propagation paths, calculated as a function of frequency, are presented in Figs. 7, 8, and 9, which include subfigures (a) and (b).

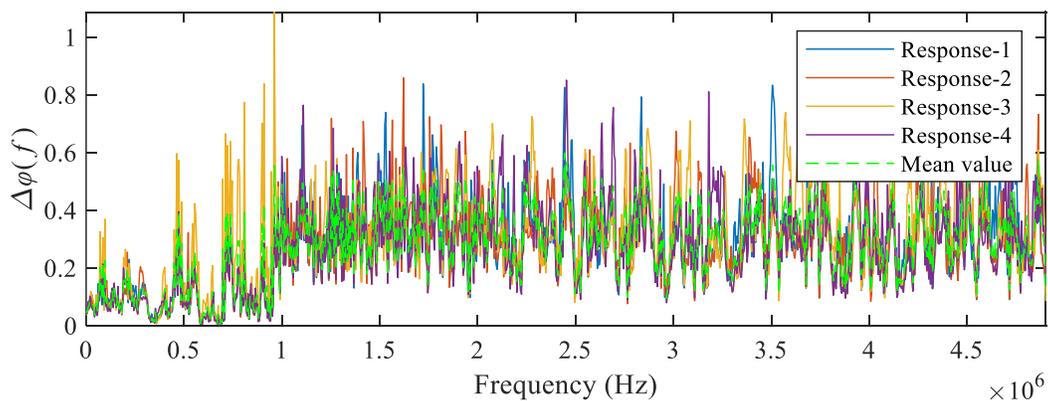

(a) Phase shift

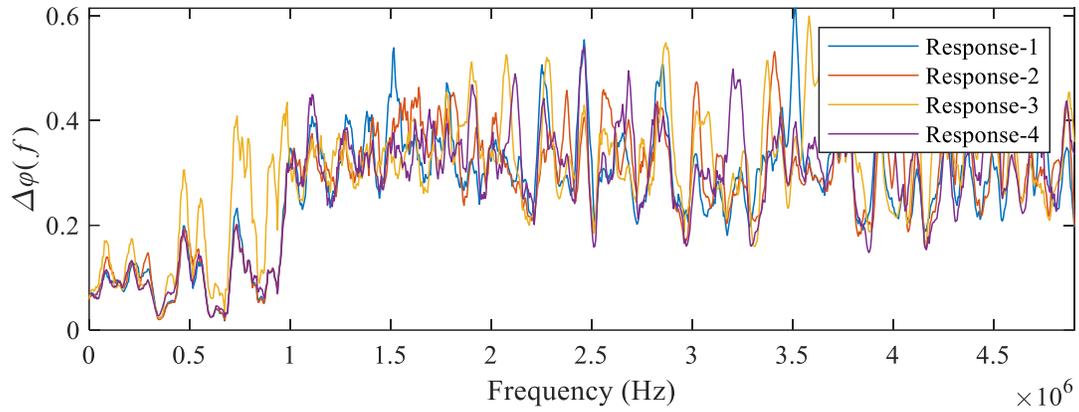

(a) Smooth trended phase shift

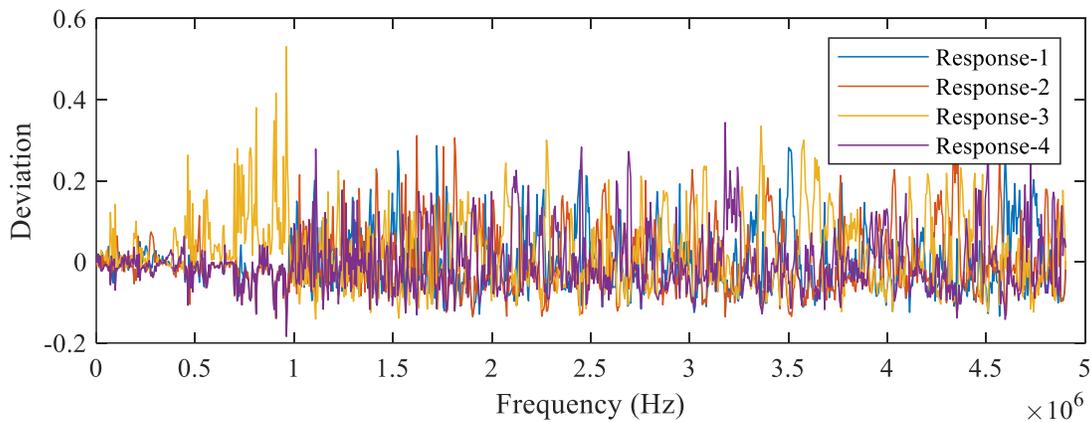

(c) Deviation from the mean

Fig. 10. Total phase shift and deviation: (a) Phase shift; (b) Smooth trended phase shift; and (c) Deviation from the mean.

In each figure, subfigure (a) displays the phase shift values, illustrating the variations in phase without any smoothing trend applied, while subfigure (b) presents the smoothed phase shift trend, highlighting the overall trend of the phase shift over time. Subfigure (c) in each of these plots shows the phase shift components along with their deviation from the mean phase shift, indicating the differences from the average value. These plots facilitate a deeper analysis of the characteristics of phase shifts as they vary with frequency. The curves below 1 MHz clearly show that the phase shift of the signal received by sensor 3, which is closer to the defect, is more noticeable, as indicated by the yellow curve in these figures.

It is evident that sensor-3, which is closest to the damage, shows significant differences in phase shift components, regardless of the propagation path, compared to the phase shifts calculated from the stronger signals received by the other sensors. This indicates that the distance between the sensor and the damage has a substantial impact on the phase shift. Meanwhile, the deviation curve clearly indicates that the phase shift components received by sensor-3 show the largest difference from the mean. This suggests that the proximity of sensor-3 to the damage results in a more pronounced variation in its phase shift. By utilizing the segmented signals collected at each sensor corresponding to different propagation paths, the total phase shift can be calculated using Eq. 2, as shown in Fig.10. It is evident that the total phase shift for sensor-3 differs significantly from the curves of the other sensors, indicating that its proximity to the damage has a particularly pronounced effect on the total phase shift.

## 5. Imaging algorithm verification and comparison

### 5.1 Algorithm verification

Once the phase shift components or total phase shifts are obtained, the geometric phase shift index (GPS-I) can be calculated using Eq. 3. Fig. 11 shows the damage index calculated using phase shift components, which are derived from the phase shift components obtained from each of the three wave packets. It can be observed that using phase shift components can also provide an approximate localization of the damage, although the error is relatively large. However, Fig. 12 illustrates the damage index distribution obtained from the total phase shift calculated using three different wave packets corresponding to the first three propagation paths. It can be seen that utilizing the total phase shift gives a more accurate localization of the damage, indicating the effectiveness of the total phase shift in damage detection.

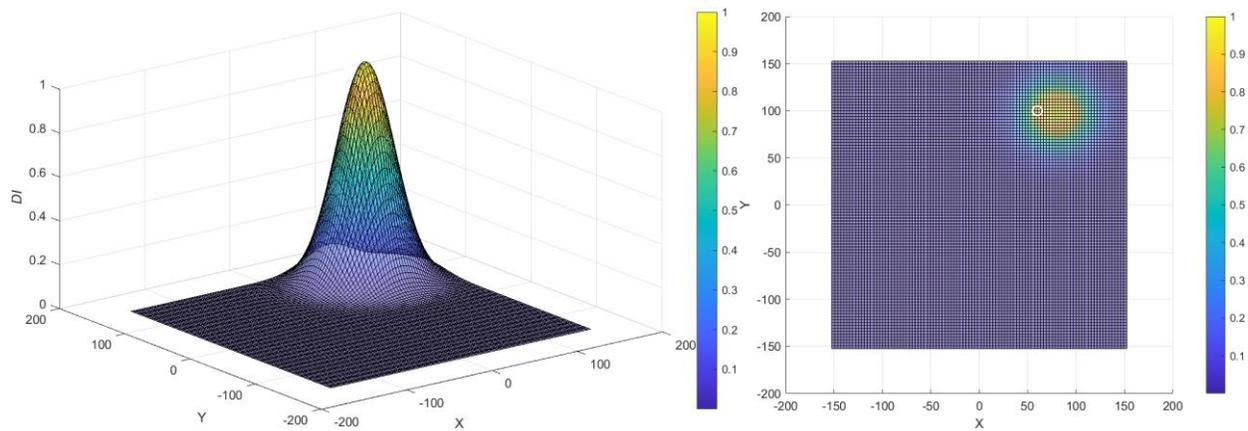

(a) 1st wave package

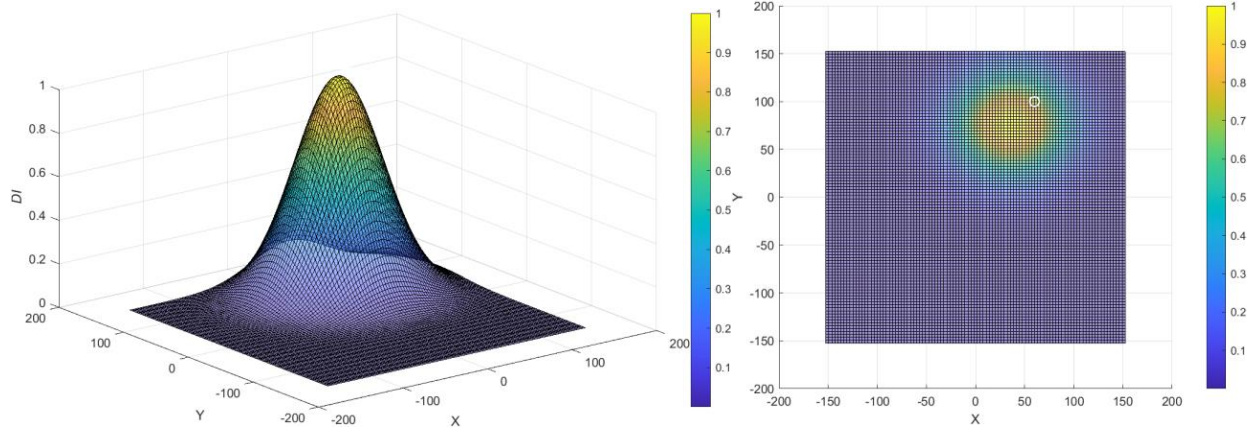

(b) 2nd wave package

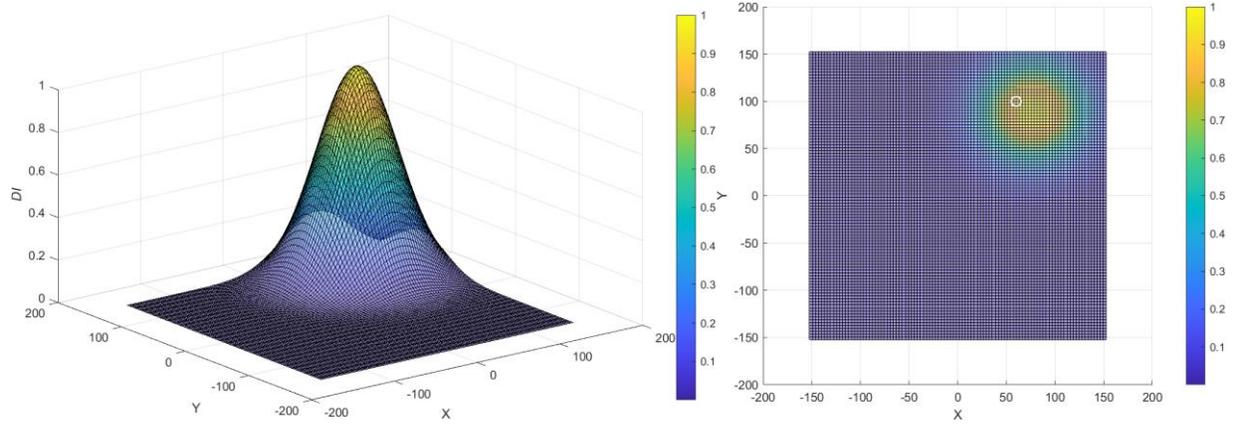

(c) 3rd wave package

Fig. 11. Damage index calculated using phase shift component based on: (a) 1st wave packet; (b) 2nd wave packet; and (c) 3rd wave packet.

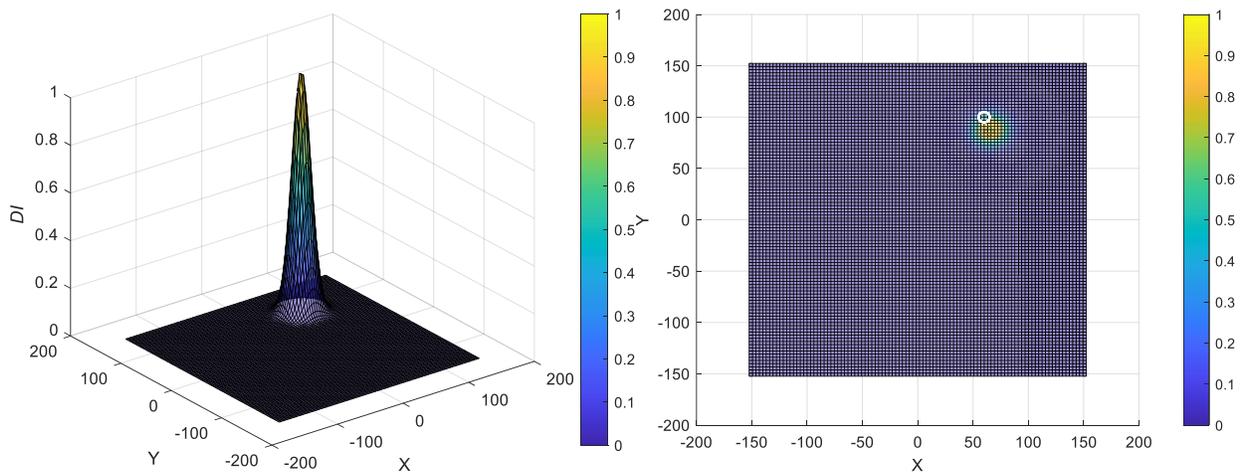

Fig. 12. Damage index calculated using total phase shift for the plate.

## 5.2 Influence of defect location

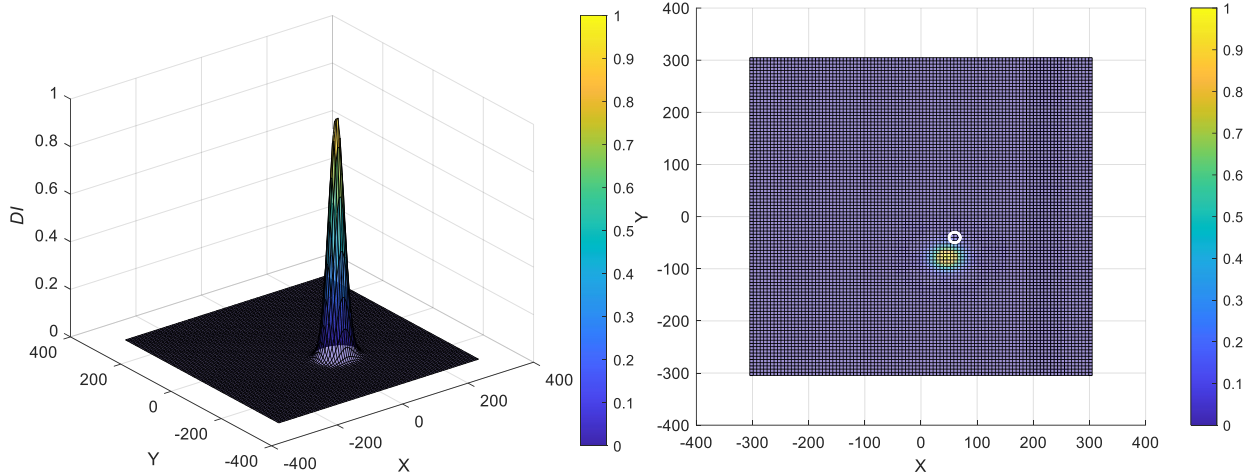

Fig. 16. Damage imaging results influenced by defect location.

Another case was considered where the damage was placed near the center of the plate. The defect is also a circular hole with a diameter of 6 mm, but it is located at the coordinates (60, -40). The final damage localization results are presented in Fig. 16. These figures show the distribution of the damage index on the plate, calculated using the total phase shift. It can be seen that the proposed method works equally well irrespective of whether the damage location is near the boundary or near the center of the plate.

## 5.3 Comparison with established damage image technique

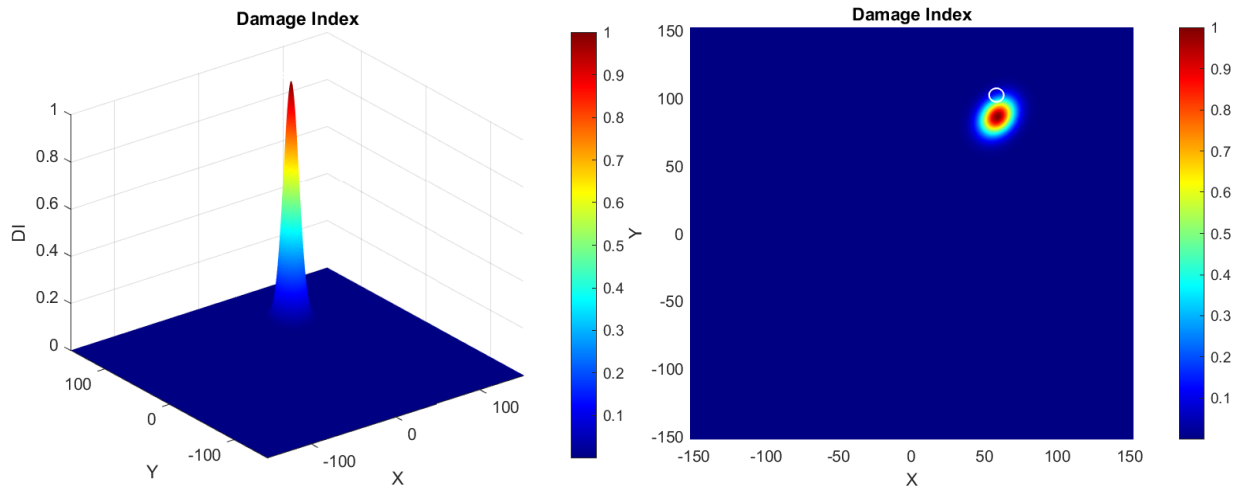

Fig. 13. Damage imaging results based on the sideband peak count (SPC) technique.

As mentioned in the introduction, we previously proposed a diagnostic damage imaging method for damage identification and localization in plate-like structures based on the SPC-I technique.

This method utilizes the SPC-I values extracted from Lamb wave response signals to create a damage map at each point on the plate, which highlights the location and approximate extent of the damage through the distribution of the calculated values. Fig. 13 presents the damage imaging results for damage detection and localization in plate-like structures using the SPC technique. By comparison, it can be observed that both the diagnostic imaging results based on the SPC-I technique and those based on the proposed TA sensing technique are quite similar, and both are capable of accurately capturing the location of the damage. However, it is important to note that the damage imaging method based on the SPC-I technique utilizes eight sensors, which are placed at the four corner points of the plate and at the midpoints of each edge, while the method proposed in this study uses only four sensors.

## 6. Conclusions

This study proposes a diagnostic imaging method for damage identification and localization in plate-like structures. The method involves collecting response signals from the four corners of the plate, segmenting the signals, and mapping them to virtual acquisition points to construct a multi-sensor Hilbert space. By calculating the phase shifts, including both the phase shift components and the total phase shift, the geometric phase shift index (GPS-I) is derived. This ultimately results in a damage index (DI) distribution map, which effectively indicates the location and approximate extent of the damage, providing a reliable basis for damage diagnosis.

Despite the promising results demonstrated in this study, several limitations must be acknowledged due to the current reliance on numerical simulations without experimental validation. To ensure a comprehensive understanding of the method's applicability and reliability, future research will focus on the following aspects: first, conducting experimental validation to further confirm the method's effectiveness; second, performing a thorough sensitivity analysis, examining the impact of parameter variations, including excitation frequency, as well as signal noise in realistic environments; furthermore, exploring the influence of different boundary conditions on damage localization accuracy and the method's applicability to materials with varying properties and heterogeneity; concurrently, investigating the use of a wider sensor network to enhance spatial resolution and sensitivity; and evaluating the method's scalability to larger and more complex structures, along with its computational complexity and processing time, to determine its suitability for real-time applications. Therefore, addressing these challenges in future studies will help enhance the method's potential for practical use.